# Atomic defects of the hydrogen-terminated Silicon(100)-2x1 surface imaged with STM and nc-AFM

Jeremiah Croshaw[1,3], Thomas Dienel[1,§], Taleana Huff[1,3], Robert A. Wolkow[1,2,3]*

1 Department of Physics, University of Alberta, Edmonton, Alberta, T6G 2J1
2 Nanotechnology Research Centre, National Research Council Canada, Edmonton, Alberta, T6G 2M9, Canada
3 Quantum Silicon, Inc., Edmonton, Alberta, T6G 2M9, Canada

**Abstract**

The hydrogen-terminated Silicon(100)-2x1 surface (H-Si(100)-2x1) provides a promising platform for the development of atom scale devices, with recent work showing their creation through precise desorption of surface hydrogen atoms.  While samples with relatively large areas of the hydrogen terminated 2x1 surface are routinely created using an in-situ methodology, surface defects are inevitably formed as well reducing the area available for patterning.  Here, we present a catalog of several commonly found defects of the H-Si(100)-2x1 surface.  By using a combination of scanning tunneling microscopy (STM) and non-contact atomic force microscopy (nc-AFM), we are able to extract useful information regarding the atomic and electronic structure of these defects.  This allowed for the confirmation of literature assignments of several commonly found defects, as well as proposed classification of previously unreported and unassigned defects. By better understanding the structure and origin of these defects, we make the first steps toward enabling the creation of superior surfaces ultimately leading to more consistent and reliable fabrication of atom scale devices.

**Introduction**

Novel approaches to advance integrated circuitry beyond CMOS, including reduction in power consumption and qubit-based computing, focus on atom scale structures and their reliable fabrication[1]. Hydrogen-terminated silicon (H-Si) surfaces are a versatile platform for the patterning and operation of atom scale devices. Such devices include qubits[2,3] and single electron transistors[4,5] made from atomically precise implanted donor atoms near the H-Si surface, and logic structures using silicon dangling bonds[6–8]. In many cases the structures' functional elements are comprised of few or even single atoms. At such dimensions, atomic scale defects of the surface and in the shallow subsurface region have a significant impact on device patternability and operation[9]. In order to develop suitable means to accommodate defects, whether it be by optimizing sample preparation, quantifying how defects affect device operation[9], or by using convolutional neural networks to autonomously identify defects[10,11,12], a comprehensive understanding of the many varieties of defects is needed.

Native silicon atoms at the unreconstructed (100) surface would extend two unsatisfied bonds into vacuum. To minimize the energy of the surface, each silicon atom bonds with a neighboring Si atom creating dimers, thus reducing the number of dangling bonds by half. Dimers extend into rows of dimers which run parallel along the surface. The study of Si(100) surface defects was one of the first applications of scanning probe microscopy[13]. The three observed species were identified as a missing silicon dimer, a pair of missing silicon dimers, and a missing pair of Si atoms on the same side of two dimers. Subsequently, the latter was re-assigned as a bonded H, OH pair from water contamination[14–16]. Further insights became available by non-contact atomic force microscopy (nc-AFM), separating the electronic and structural behaviour of the Si(100) surface[17].

The addition of hydrogen to surface silicon atoms saturates all available bonds[18] and leads to the formation of mainly three surface reconstructions. The 2x1 phase—most commonly used in hydrogen lithography due to its ease of *in situ* preparation of large, defect free areas[19]—has each surface Si atom in a dimer bonded to one hydrogen atom. The 1x1 phase forms when the dimer bond is broken and each surface Si atom bonds to 2 hydrogen atoms, forming a silicon dihydride ($H_2$-Si). The 3x1 phase is a combination of the previous two, containing alternating 2x1 dimers and 1x1 dihydrides[20,21]. With continued study, it became apparent that the complexity of possible surface reconstructions and surface defects extended well beyond those initially observed. Here, we provide a comprehensive analysis of the H-terminated Si(100) surface and its structural features. Six different scanning probe imaging modes are used to confirm the atomic structure of several commonly reported defects, as well as to classify previously unknown defects. The latter includes a neutral point defect that can be found to be decorated with a single H-atom, rendering the whole structure negatively charged. Finally, we demonstrate the successful removal of the weakly bound H atom through tip manipulation.

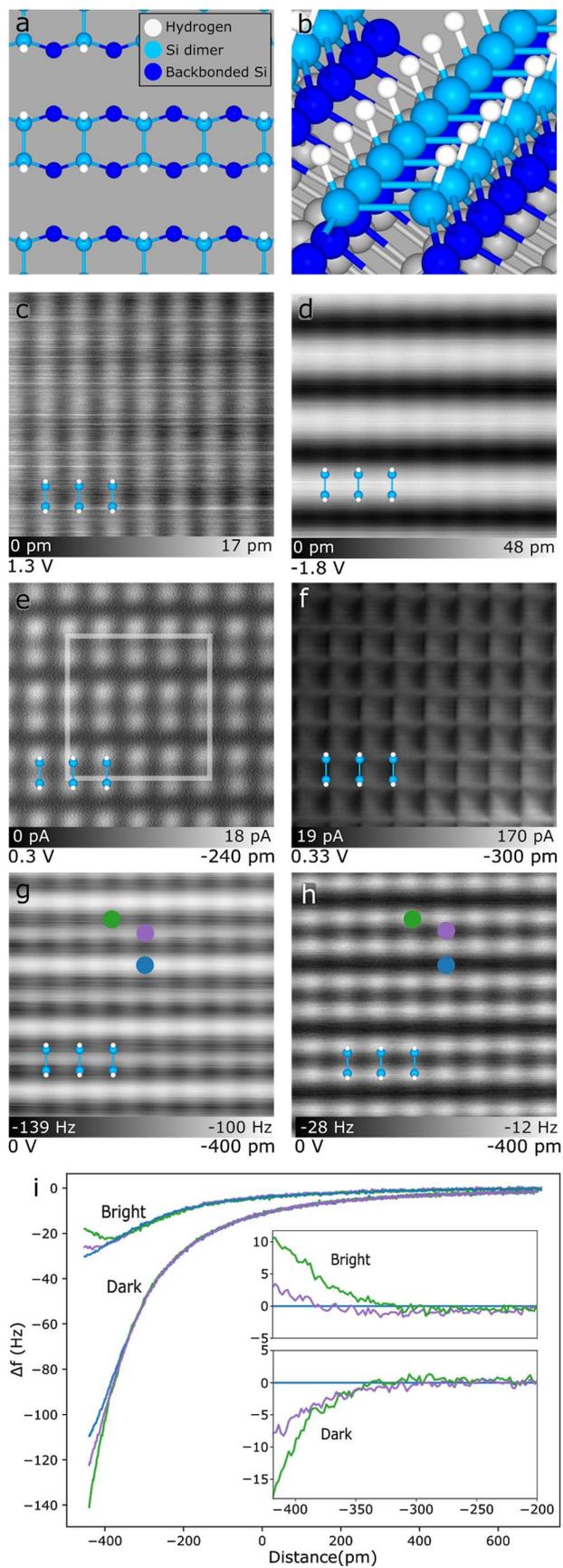

**Figure 1. Surface of H-terminated Si(100)-2x1 in different imaging modes.** (a) and (b) Top and side view of structural model, size of area shown in (a) indicated in (e). (c) and (d) Constant current (I=50pA) STM topography of empty and filled states (bias voltages as indicated). (e) to (h) Constant height images of the same area as in (c), indicated height offset Δz initialized using tunneling setpoint parameters of I = 50 pA and V = -1.8 V over a H-Si/metallic STM tip (e), STHM (f), nc-AFM frequency shift in dark contrast (g), nc-AFM frequency shift in bright contrast (h). (i) Position-dependent frequency shift spectroscopy Δf(z) in bright and dark nc-AFM mode (positions indicated in (g) and (h)). The inset shows the calculated difference spectra in reference to the spectra taken between the dimer rows (blue position marker).

**Results and Discussion**

The variability of scanning probe imaging modes originates from the applied feedback mode, different tunneling parameters, or the functionalization of the probe tip. Figure 1 demonstrates various imaging modes for the defect-free H-Si(100)-2x1 surface. Figure 1a and 1b show a ball-and stick model of the surface reconstruction with the pairing of surface Si atoms into dimers and the termination of the residual dangling bonds with hydrogen (See Methods). For completeness, we start with the well-known STM topographies of empty and filled states (constant current imaging, sample bias as indicated, Figure 1c and 1d). Imaging both states allows for the assignment of dimer rows. Recent work has found that tip functionalization can determine the contrast sharpness and apparent atomic positions in empty state STM images of the H-Si(100) surface[22]. Three examples of this effect are displayed in Supplementary Information Figure S1, highlighting the necessity to include subtle surface features in the assignment of dimer position.

Further insights can be gained from constant height images. Figure 1e shows a constant height STM image of the surface at a sample bias which probes the onset of the conduction band (donor band) of our crystal[9,17,24]. Hydrogen-terminated silicon atoms within each dimer are identifiably without any significant convolution from bulk states. Figure 1f shows the final STM imaging method used, known as scanning tunneling hydrogen microscopy (STHM)[25]. The use of a flexible species at the apex of a metallic tip to provide enhanced contrast was first reported using CO-functionalized AFM tips to image the molecular structure of pentacene[26]. Other functionalizations of AFM tips have been explored including Cu-O tips[27,28] and Xe tips[29,30]. The use of $H_2$[31,32] and $D_2$[25] was the first successful demonstration of the enhanced imaging contrast using STHM. Rather than direct tip functionalization as done in AFM, STHM was routinely achieved by leaking in a background of molecular hydrogen (~$10^{-9}$ Torr) until an $H_2$ molecule becomes trapped in the tip-sample junction. We are able to achieve STHM resolution by directly functionalizing the tip apex with hydrogen using voltage pulses over the H-Si surface, as discussed elsewhere[22,33,34]. Our ability to achieve STHM resolution using an H-functionalized tip aligns with recent theory that suggests the $H_2$ molecule actually dissociates resulting in an H-functionalized tip [35]. In STHM imaging, the surface appears as a series of squares with each intersection correlating to a H-Si atom. The image's slight asymmetry is commonly attributed to the shape of the tip apex and the location of the H atom. Figure S3 shows a variety of images of the H-Si(100)-2x1 surface acquired with a variety of flexible tips.

A true force feedback can be visualized with frequency-shift maps generated by non-contact AFM. Based on the apparent contrast of the hydrogen atoms with the surrounding surface we denote the two different modes as bright and dark contrast AFM. Previous studies have identified these two contrasts when imaging the hydrogen-terminated Si(100)[36] surface, where the dark contrast image corresponds to a Si-terminated tip[37] and the bright contrast image corresponds to an H-terminated tip[37]. Figure 1i shows the height-dependent frequency shift $\Delta f(z)$ that is governed by the interaction between tip and sample at selected positions on the surface. To highlight the termination-dependent reactivity we calculate the difference in frequency shift[38] compared to the "between dimer" position, where the least interaction is expected. The strongest repulsive component is clearly observed for the bright contrast tip on the H-atom, while the Si-terminated tip leads to the strongest attractive interaction at the same site (dark contrast).

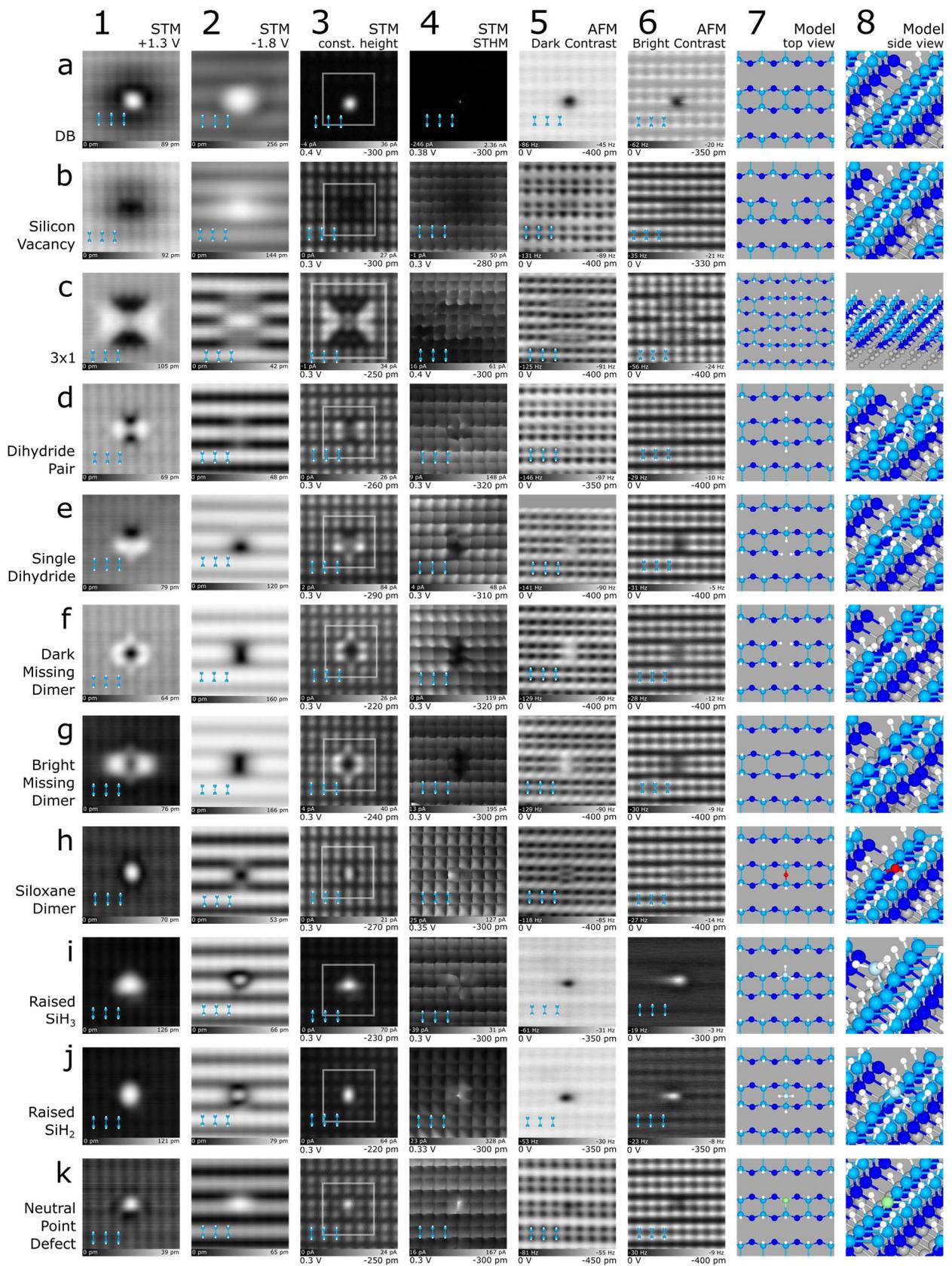

Figure 2. **Common features of the H-Si(100)-2x1 surface imaged using the indicated STM and nc-AFM imaging modes.** The dimer structure is shown as an overlay in each scan (lower left corners). The area of the structural models in columns 7 is indicated by the grey box surrounding the defect in column 3. Colours in the ball and stick models are: H - white, Si dimers - light blue, back-bonded Si - dark blue, O - red, raised Si - lightest blue, and unknown species - green.

Combining all six imaging modes allows for an in-depth characterization of the most common surface features and their local environment as shown in Fig. 2. While not an exhaustive list, it features defects routinely seen when the surface is prepared using the procedure outlined in the Methods section. To categorize the defects, one could focus on functional aspects (charged vs. neutral features)[9], or structural aspects including missing or additional atoms. Here, we categorize the defects solely on structural commonalities, based on the number of affected atoms.

First, we list features that involve only one side of a dimer. These include a missing H atom that leaves the dangling bond (DB) of the underlying Si atom exposed (Fig 2a), a subsurface vacancy (Fig. 2b), the substitution of an H atom with $SiH_3$ (Fig. 2i), and an unidentified surface point defect (Fig. 2k). Second, defects affecting a whole dimer are: two missing H atoms creating a bare dimer (neutral, shown elsewhere[39,40]), two additional H atoms on a dimer to create a dihydride pair (Fig. 2d), a single added H atom forming a dihydride on one side of the dimer(Fig 2e) that in turn requires the neighboring Si atom to be missing (Fig 2e), the removal of the whole dimer (two missing Si atoms) accommodated by H-termination or dimer formation in the second layer (dark or bright, Figs. 2f and 2g). Alternatively, the dimer bond could be replaced by either a $SiH_2$ group (Fig. 2j) or an oxygen atom (siloxane bridge, Fig. 2h). Remaining larger features can be described as a combination of two or more of the smaller structures, like the 3x1 reconstruction which is a combination of dimer rows and dihydride atoms (Fig. 2c).

The most obvious deviation from the hydrogen-terminated surface are silicon DBs, well-understood un-terminated silicon atoms[41–43]. In STM, the centre of a DB is imaged as a bright protrusion indicating the highly conductive orbital which extends into vacuum. Due to the degenerate doping of our substrate (see Methods)[6,42], DBs are usually negatively charged and the associated band bending leads to a halo around the DB in empty states imaging (Fig. 2a-1)[42,44]. The STM filled states image lacks the charge induced band bending around the DB due to competing electron emptying and filling rates rendering the DB neutral on average[43,44]. The constant height STM and STHM images in Figs. 2a-3 and 2a-4, also reveal the central protrusion but in addition, the STHM image shows a significant increase in tunneling current over a DB (36pA vs. 2.4nA). The reason is that the different tip functionalizations result in different tip-sample separations when the tip height is initialized using the tunneling setpoint (-1.8 V, 50 pA) as a fixed reference (direct comparison shown in Figure S2). Finally, AFM analysis in Figs. 2a-5 and 2a-6, presents the DB as a large negative frequency shift due to the strong interaction for both tip terminations (further details of all dark contrast images in column 5 are discussed with Figure 3).

Figure 2b shows a suspected silicon vacancy (discussed in more detail in Figure 4) previously referred to as a type 2 (T2) defect[45]. Prior works have suggested this defect originates from many sources including a negatively charged As dopant[45], Si-vacancy hydrogen complexes[9], and B dopants[46,47]. Various types of crystal vacancies have previously been identified using scanning probe microscopy including Ga vacancies in GaAs[48], As vacancies in GaAs[49], and P vacancies in InP(110)[50] which exhibit similar imaging characteristics to the Si vacancy. From the STM empty states image in Figure 2b-1, it can be seen that the vacancy exists as a negatively charged species as evidenced by the reduction in brightness from charge induced band bending. Additionally, it lacks the bright centre found with the DB suggesting the defect does not contain any orbitals which extend into vacuum. This is further corroborated by the filled states STM image in Figure 2b-2 showing a diffuse bright protrusion similar to the DB, but of reduced conductivity. In 2b-3, a reduction of current due to charge induced band bending can be observed, along with a distortion of the density of states of the surface H-Si atoms above the sub-surface vacancy defect. 2b-5 and 2b-6 show similar distortions in the AFM frequency shift signal around the two H-Si atoms closest to the vacancy, suggesting a modification in their position or reactivity.

A silicon adatom replacing an H atom (on top of one of the dimer atoms) will be fully saturated and form a raised $SiH_3$ or silyl group (Fig. 2i)[51]. Larger clusters of these Si groups can subsequently bond and form islands[21,52]. The triangular shape of the defect as seen in the STM images demonstrates the tetrahedral bonding orientation at the Si, with the forth bond affixing it to the side of the dimer beneath. The raised nature of the silyl group dominates the current in constant height imaging which leads to extended distortions in STHM (Fig. 2i-4), partially due to the groups high flexibility, and gives rise to stronger frequency shifts in nc-AFM (Fig. 2i-5,6).

Instead of bonding with the neighbouring Si atom creating a dimer bond, the Si atom could bond with 2 H atoms, whereby a dihydride is created (Fig. 2e). Although not formally reported, it has been observed in the past (such as the top right of Figure 2 in Ref. [[53]]). The remaining Si atom of the dimer is either absent (Fig. 2e) or also bonds with 2 H atoms creating a pair of dihydrides (Fig 2d)[20,53–55]. Based on their appearance, the latter defect is commonly referred to as the "bow-tie" defect[56] or simply a dihydride. The concentration of dihydrides can be controlled by the annealing temperature during sample preparation[18,57].

The $H_2$-Si units appear as depressions in empty state topography (Figs. 2d-1 and 2e-1) and correspondingly as sites of lower current in the constant height imaging (Figs. 2d-3 and 2e-3), whereas neighboring dimers appear enhanced[51]. The concomitance of a depression and enhanced surrounding atoms in empty states imaging is a similarity that dihydrides share with missing dimers (Figs. 2f and 2g)[13,58,59]. The filled states imaging provides distinction with almost negligible variations of the dihydride sites (Fig 2d-2) compared to the distinct depression of missing atoms (Figs. 2e-2, 2f-2, and 2g-2). The STHM image of figure 2c-4 allows for a direct comparison between the dihydride and the missing atom.

Above the dihydride we see a significant deviation from the expected square intersections above the H$_2$-Si atom, suggesting that multiple atoms are present. The missing atom portion of Figure 2c-4 shows an absence of the expected square intersections confirming the absence of an atom. These effects are confirmed in the dark mode AFM image (Fig. 2c-5) with the dihydride showing a splitting of the Si atom corresponding to each of the two H atoms while the missing atom shows a reduction in frequency shift characteristic of an increased tip-sample separation. The splitting of the dihydride site is not readily apparent in the bright contrast AFM image, likely due to the flexibility of the H-sensitized tip

The two variations of the missing dimer appear almost identical, except when comparing the relative height of the neighbouring dimers in constant current images of 2f-1 (dark) and 2g-1 (bright). The data were taken sequentially while continuously monitoring for tip changes, ensuring identical apex character. Similar to the unterminated surface[60], the two varieties originate from different terminations of the exposed back-bonded silicon atoms. The formation of two dimer bonds—orthogonal compared to the top layer dimers (model in Fig. 2g-7)—corresponds to the bright variation (2g) and H-termination of all sites leads to the dark variation of the missing dimer (2f) . The latter has been reported to cause less lattice strain[61], matching our observation of less distortions in the STHM and nc-AFM images near the dark missing dimer (Figs. 2f-4,5,6 compared to Figs. 2g-4,5,6). This family of defects (dihydrides, missing dimers and combinations thereof) underlines the importance of combining several modes of SPM imaging to unambiguously assign the present structure.

Figure 2c explores a structure involving a larger number of dihydrides. During H-termination, a set of two silicon atoms will bond so that the center two atoms create a single dimer row (here 3 dimers long) leaving two lone Si atoms at either side which bond with two H atoms creating dihydrides. This is the motif of the 3x1 reconstruction shown here as a three unit patch (See Figure 2c-7)[19,62,63]. Figures 2c-1 and 2c-3 show the enhancement in current at dimers neighbouring the reconstruction and a reduction above the H$_2$-Si atoms. Figure 2c-2 shows a dimer has been formed between two of the regular dimer rows, with Figure 2c-3 also highlighting this realignment. The dihydride features in the 3x1 reconstruction show similar features to the single dihydride structure mentioned above including the deviation in the STHM image (Figure 2c-4) and the splitting feature of the AFM image in Figure 2c-5. A direct comparison of the differences in nc-AFM dark mode imaging including the identification of individual H atoms at the dihydride sites is given in Figure 3.

Defects that we occasionally observe as substitutions that replace the dimer bond manifest themselves through the insertion of additional atoms between the dimer bond. We representatively show a SiH$_2$ group and an oxygen atom in such a position. The tetragonal bond orientation of the silicon atoms forces both substitutions into a raised height configuration (Figs. 2i-7 and 2j-7) compared to a regular dimer. The former represents a silicon adatom in the bridge position between two top layer silicon atoms compared to the already discussed SiH$_3$ group bonded to one side of the dimer. Both adatoms show similar apparent height in STM topography with their location with respect to the dimer easily discerned. In dark contrast AFM images the chemically reactive silicon tip strongly interacts with both defects (Figs. 2i-5 and 2j-5). Conversely, the bright contrast images in 2i-6 and 2j-6 are governed by the partially repulsive interaction with a hydrogen-terminated tip. Both defects are raised above the surface to account for the observed frequency shift over the defect location compared to the background H-Si(100) surface. Here, AFM does not allow one to discern the groups of individual hydrogen atoms due to the flexible nature of the raised clusters, creating the distorted appearance when pushed in constant height imaging.

The bridging oxygen in Figure 2h forms a siloxane dimer (see structure in Fig 2h-7), sometimes described as a split dimer[64,65] or incorrectly identified as a dihydride[2,66]. STM imaging reveals a very localized defect with only subtle impact on neighboring dimers (Figs. 2h-1,2,3). Figure 2h-4 shows slight variation from the 2x1 dimer, with the positions of the atoms in the dimer further apart with a central bright feature present. This agrees with the AFM data in Figure 2h-5, showing the dimer contains a third atom in the middle. In bright mode AFM the attractive interaction between H-tip and the oxygen's non-bonding electron pairs leads to a slight depression in the centre and lets the dimer's H atoms appear pushed out (Fig. 2h-6). A defect of similar appearance was found on chlorine-terminated silicon[67], and was associated with water contamination in the vacuum chamber (observed as H and OH bonded to the unterminated surface[15,68,69]) where a mild annealing followed by halogen termination allowed the oxygen to enter into the dimer bond. The authors comment that they expect this feature to also be present on hydrogen-terminated silicon.

Figure 2k shows an unknown neutral point defect. This defect is also observed in a negatively charged, H-decorated state. By removing the hydrogen from the structure via tip interactions (as discussed with Figs. 5, S5, and S6), the species transitions to the more stable, H-free, neutral variety. The observation that the H liberation occurs with tip interaction makes it difficult to obtain a full set of images of the H-decorated variety. The neutral variation exists very close to the surface both spatially and electronically as indicated by the height scales in the constant current images in Figure 2k-1 and 2k-2. It shows only a slight increase in conductivity in the constant height image of Figure 2k-3 localized to a single atomic site, suggesting it may exist in place of the Si atom on one side of the dimer. The AFM image of Figure 2k-5 confirms this localized nature, with the defect appearing as a slightly darker circular feature. Figure 2k-6 similarly shows a slightly darker feature, indicating additional reactivity with a hydrogen-terminated apex, but does not provide any further information on the structure.

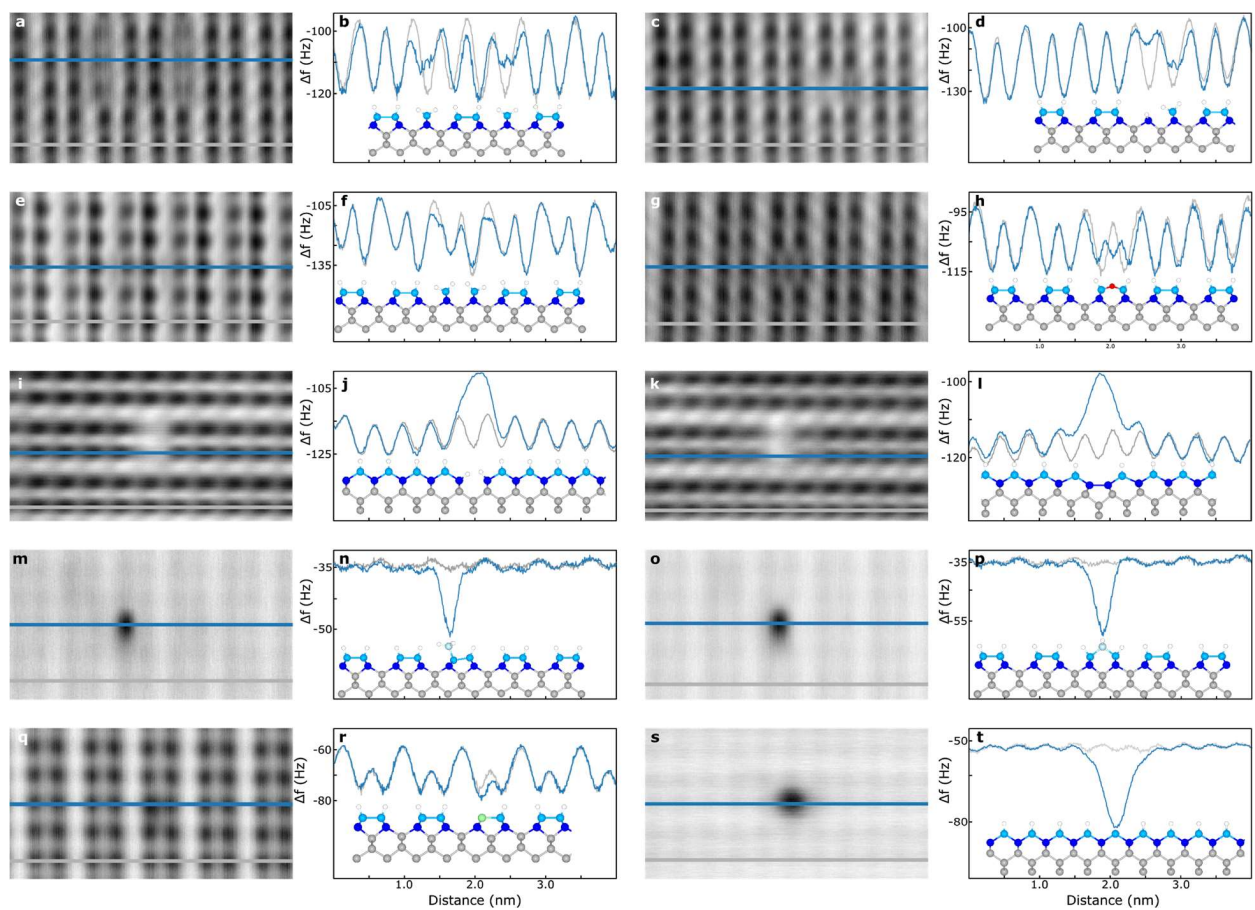

Figure 3. **Analysis of defects using dark-contrast AFM and line scan analysis.** Line cuts of the dark contrast AFM scans were performed by averaging over 6 adjacent lines, with the thickness of the line profile indicating this averaging. The defect profiles are indicated in blue, while the defect free 2x1 phase sample for each panel is in gray. The defects explored were (a,b) 3x1, (c,d) a single dihydride, (e,f) a dihydride pair, (g,h) a siloxane dimer, (i,j) a dark missing dimer, (k,l) a bright missing dimer, (m,n) a single raised $SiH_3$, (o,p) a single raised $SiH_2$, (q,r) a neutral point defect, and (s,t) a DB. See Figure 2 for the experimental image parameters.

The dark contrast AFM achieved the most striking resolution of all imaging modes presented in Fig. 2 presumably due to the Si-tip's greater chemical reactivity and apparent lack of apex flexibility. In the following we compare line profiles across the presented defect structures and a defect-free region with a simple structure model optimized by molecular dynamics calculations (Merck molecular force field 94, see Methods). Starting with dihydride based defects, Figure 3a,b shows an analysis of the 3x1 reconstruction. In Figure 3b the defect line profile (blue) is superimposed on a normal 2x1 phase line profile (gray), highlighting the reported splitting over the two dihydride-hosting atoms. The two central

Si atoms form a positionally shifted dimer in between the dihydrides, with the formed dimer bond reducing the distance between the constituent Si atoms  The 3x1 example can be compared to the case of a single dihydride in Figure 3c,d, showing a similar split over the location of the dihydride.  A slight depression is seen above the missing Si atom but of reduced frequency shift when compared to the defect-free cross section, suggesting the tip cannot fully probe the vacancy due to the close proximity of the dihydride atoms. Examining the dihydride pair of Figure 3e,f, it interestingly appears very similar to the normal 2x1 phase with only slight variation at the position of the outermost H atoms.  The lack of a hydrogen splitting feature, as with the other dihydride species, can be explained by looking at the calculated geometry of the two closely-spaced dihydride species in the dimer.  Local repulsion between the two nearest hydrogen atoms results in a rotation of the dihydrides with respect to the surface.  The position of the outer two hydrogens are now further away and much closer to the bulk atoms, making the feature less visible to the AFM tip.

Examining the siloxane dimer in Figure 3g,h shows three distinct minima in place of the dimer.  The outer two peaks are spatially extended compared to a regular dimer due to the need to accommodate the oxygen atom bonded between the two Si atoms.

Figures 3i,j probe the dark missing dimer, with the hydrogen terminated second layer silicon atoms at the defect location presenting as less attractive to AFM probing due to increased distance (a frequency shift maxima in the middle).  Figures 3k,l show the bright missing dimer. Examining the defect location in Figure 3l, two local minima "shoulders" are apparent on the sides of the central peak.  These have a less negative frequency shift, with the minima pulled spatially towards the centre of the defect. This observation is a result of the second layer Si atoms bonding together at the defect location, introducing horizontal lattice strain. This pulls the neighboring atoms inwards more than in the dark missing dimer case, accounting for the shoulders in 3l which are absent in 3j.  Importantly, these two defects were imaged in succession so any variations between the two are a consequence of their differing nature, not a tip change.

The two raised Si species in Figures 3m,n,o,p present an extra challenge to analyze because they must be imaged with a larger tip-sample separation to prevent tip-defect contact.  As a result, the frequency shift of the surface is of smaller magnitude compared to the defect signal (compare blue and grey line traces in Figure 3n,p).  Despite Figure 3n,p showing the defect as a mostly featureless peak in both cases, the position of the defect with respect to the lattice can still be extracted;  comparing the two panels it can be confirmed that the $SiH_3$ originates above one side of a dimer while the $SiH_2$ is centered between the two atoms of a dimer.  Additionally, the singly bonded $SiH_3$ is a narrower peak (blue cross section of Figure 3n) than the doubly bonded $SiH_2$ one (blue line Figure 3p), perhaps due to the extra degrees of freedom allowing the defect to bend more upon AFM examination.

The neutral point defect in Figure 3q,r displays as a slight decrease in the minima above the defect, suggesting a similarly coordinated substitutional species of enhanced electronegativity compared to a H-Si atom.

Lastly, a lone DB is shown in Figure 3s,t. As expected, the DB shows up as a localized feature of enhanced reactivity when compared to the surface.  This reactivity extends spatially away from the DB location, generating a broad minimum that eclipses the signal from neighboring lattice sites.

A more in-depth analysis of the silicon vacancy defect from Figure 2b is now given. While all of the other examined defects are observed with the same consistent appearance, the Si vacancy has been found to vary. Figure 4 shows empty states, filled states, and constant current images of 4 different configurations of the defect.  Each of the configurations in 4a-d, e-h, and i-l correspond to a silicon vacancy, but located at different lattice sites as shown in the provided ball and stick models.  Starting with Figure 4a (labelled I), the defect is centered around a surface lattice site affecting one side of a Si dimer, with the negative charge of the species bending the bands down locally, as shown by the radial dark depression around the defect center. The filled states STM image of 4b shows an increase in measured height, correlating again to the fixed negative charge of the defect[45].  Looking at the constant height STM image in Figure 4c, it appears that no atom is present at the defect lattice site.  This is supported through examination of the AFM line cuts of this defect in Figure 4q,r (and Figure S4). A more negative frequency shift minimum is seen in the remaining atom of the dimer, with a shift of the minimum toward the site of the vacancy as shown in 4r. The vacancy itself shows a smaller minimum above its location (blue curve in 4r), with the magnitude of this minimum comparable to measurements over equivalent second layer Si atoms as would be measured in a cross section taken between dimers (burgundy line in 4q).  This observation suggests there is no atom present, leading to a classification of this species as a Si vacancy[70,71]. These are predicted to exist as negative charge centers due to the degenerate doping of the crystal (See Methods).  Figure 4e,h shows another silicon vacancy (labelled II), but now positioned at a different lattice site one atomic layer below the surface (as referenced to the Si atoms of the dimers).  While the broad features of the STM probing in Figure 4e-g are similar, it can be seen that the defect no longer appears to affect a single atomic site, but rather reduces the apparent height of two adjacent dimers as shown in 4g. While the subsurface defect cannot be directly probed, the similarities it shares with Figure 4a strongly supports its assignment as a Si vacancy below the surface (See the ball and stick model in Figure 4t). This is further corroborated by line-cut analysis in Figure 4s,t, where the two atoms centered above the defect show a reduced minimum, as well as a horizontal shift in position towards the defect center due to a polaronic distortion induced by the vacancy's localized negative charge. The ball and stick models of Figure 4 h and t have been manually edited to show this effect.

 Figure 4i-l shows the third common type of Si vacancy (labelled III).  This defect shares broadly similar STM features when compared to the previous vacancy examples  but now spatially shifted with respect to the Si lattice; the dark depression in Figure 4i is now symmetric in appearance and centered in the dimer row, the bright enhancement in 4j extends over more dimers, and in the constant height STM in 4k the two dimers above the defect center show a reduction in apparent height.  Aligning the centre of this defect to a model of the underlying Si lattice (See Figure 4l), along with the symmetry observed in STM analysis, it is likely the position of this vacancy is in the third atomic layer.

Figure 4m-p looks at a possible final silicon vacancy configuration (labelled Ia). With the addition of hydrogen to the chamber during sample preparation (See Methods), it is suspected that some of the hydrogen radicals can penetrate a few monolayers into the surface, allowing them to passivate dangling bonds from Si atoms immediately surrounding a vacancy[70,71]. Evidence of this is suggested by the vacancy in 4m being similar to the one shown in 4a, but with a few key differences. To first order, both display a dark depression on one side of a dimer indicating a missing silicon atom. Differences become apparent however when examining adjacent lattice sites to the missing atom. In 4a, the surface hydrogen atoms around the vacancy location have similar character, displaying as darker than the surrounding surface. This is in contrast to 4m where H-Si atoms to the left of the vacancy appear unperturbed, and hydrogen atoms below and to the right are of differing character. Further differences in the appearance of this configuration are also seen in the STM image of 4n when compared to 4b, with 4b having two maxima and 4n only one. The normal appearance of the surface immediately to the left in 4m supports the idea of a hydrogen penetrating the surface satisfying the left-most sub-surface dangling bond associated with vacancy (See Figure 4p), reducing the lattice strain and making the surface hydrogen above it better match the defect-free surface. Figure 4o shows an STHM image of the vacancy. This imaging method was used instead of a constant height image due to the tip functionalization at the time of imaging. Similar features consistent with the interpretation of the vacancy can be observed, mainly, the absence of corners in the STHM image above the vacancy as well as a slight negative shift in the current above the lattice intersections corresponding to the right and lower H-Si atoms.

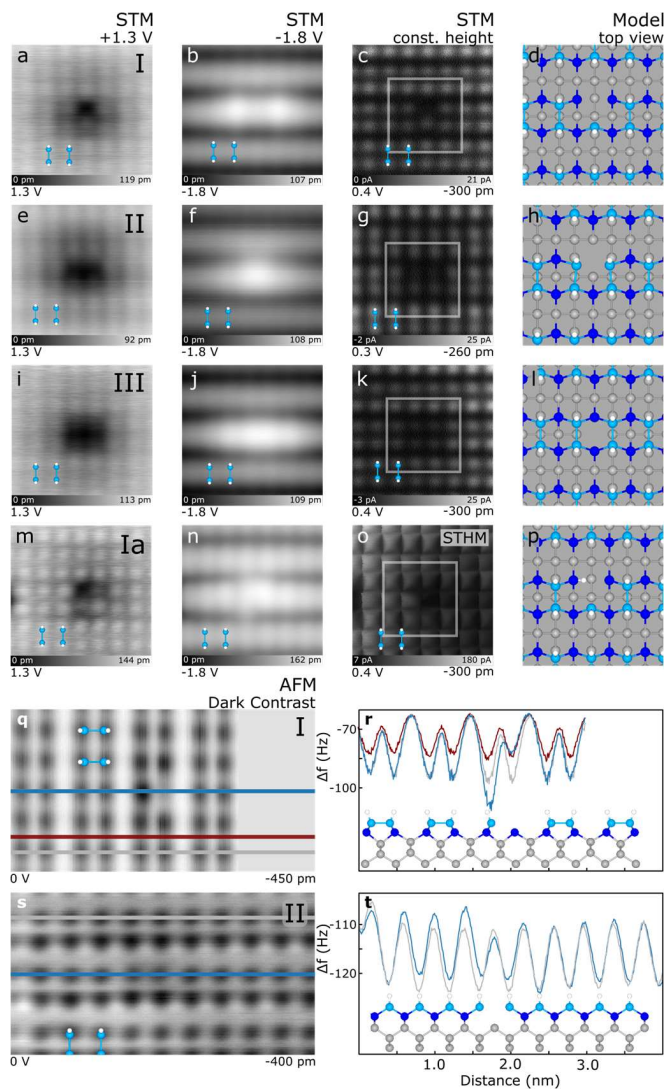

Figure 4. **Silicon vacancies at different lattice sites.** (a-d), (e-h), and (i-l) show the changing STM appearance of silicon vacancies located at different depths and lattice sites, labeled I, II, and III, respectively. (m-p) shows a suspected variation (Ia) of (a-d), in which one of the dangling bonds left from the vacancy has possibly bonded to a hydrogen. Each STM image is 3.4 nm x 3.4 nm. (q,r) show the linecuts of vacancy I in (a) while (s,t) show the linecuts of vacancy II in (e).

We now return to a discussion of the origin of the neutral point defect. While its true nature remains inconclusive, Figure 5 provides three case examples showing they can weakly hold a H atom rendering the entire structure negative. The negatively charged nature of the H-decorated point defect can be seen in the STM empty states images of Figure 5a,d,g. In each case, the negative charge associated with the H atom is indicated by the charge-induced band bending surrounding its location. Upon scanning at a filled states imaging bias (Figure 5b,e) or using AFM (Figure 5h), a sharp change in contrast is seen when scanning over the defect  Subsequent imaging of the defect after this change in contrast is given in the empty states images of Figure 5c,f,i, showing the H atom has been removed from the structure leaving the neutral point defect with charge-induced band bending no longer present. Further examples of the H removal along with STM I(V) and dI/dV spectroscopy of the neutral point defect can be seen in Figure S5 and S6.

Our assignment of the neutral point defect as a hydrogen trap is motivated by two observations, the first indication being the change in contrast in the AFM panel of Figure 5h. As discussed earlier, dark contrast AFM is from a silicon-terminated apex and bright contrast from a hydrogen-terminated one. The change between these two contrasts at the middle of the frame suggests the tip has potentially picked up a hydrogen atom from the defect, changing the AFM imaging character. The second observation supporting a hydrogen trap is the similarity of the negative state of the defect to physisorbed hydrogen atoms on the surface. Prior work reported that hydrogen atoms on the surface are negative under degenerate n-type doping[9], while another showed they could be picked up from the surface with filled-states STM imaging[33]. Their similarity in appearance can also be observed in Figure S7. This evidence, together with the fact our sample preparation methodology produces many hydrogen radicals that can penetrate the surface, supports the idea that the neutral point defect behaves as a hydrogen trap. It has been reported in the literature that boron, when added to silicon, can behave as a hydrogen trap [72–75]. However, the areal concentration these neutral point defects are observed at (0.8-7.6 defects/10 nm$^2$) is higher than would be expected for our arsenic doping level, or for contaminant boron from the commercial wafer processing[45]. Further investigation into the origin of this defect will be required for a conclusive determination.

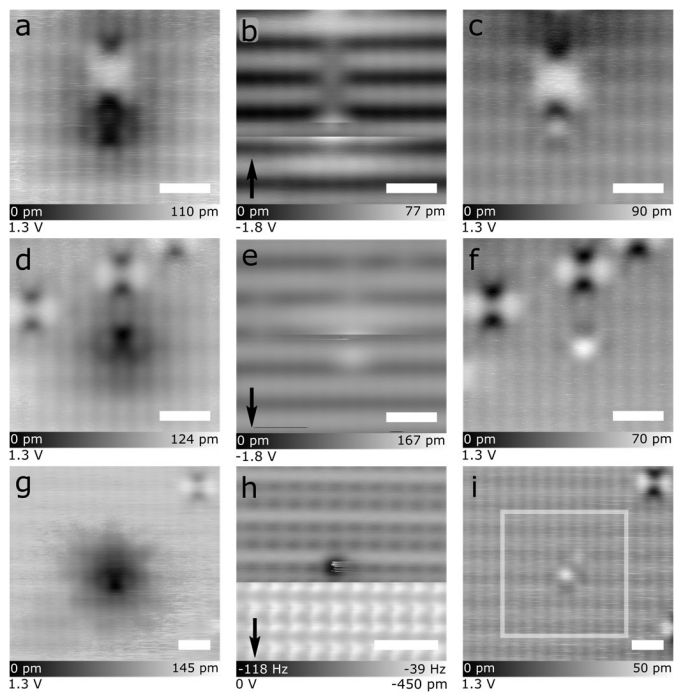

Figure 5. **Removal of a hydrogen atom from the neutral point defect.** (a-c), (d-f), and (g-i) show three instances of the neutralization of a negatively charged defect which is the precursor form of the neutral point defect. (a,d,g) Empty states images of the negatively charged H-decorated point defect before H

liberation. (b) and (e) show filled states STM removal events, while (h) shows an AFM removal event. The arrows in the lower left indicate the scan direction, with contrast changes in the scan indicating a removal event has occurred. (c,f,i) Empty states STM images of the same frames as (a,d,g), but after the H removal. The point defects no longer display charge-induced band-bending around their location. The white box in (i) indicates the size of the scan frame in (h). Scale bars are 1 nm.

**Conclusion**

In this work we have created a comprehensive list of commonly found defects on the H-Si(100)-2x1 surface analyzed using a combination of several STM and nc-AFM imaging modes. Through this analysis we are able to extract information about the electronic states of the defects, as well as probe their atomic structure using differently reactive AFM tips. As a result, we were able to confirm the classification of several surface defects, as well as more confidently identify the previously reported T2[45] defect as having character more consistent with a silicon vacancy. Finally, we examined the previously unreported neutral point defect, observing their transition from a negatively charged species through the proposed liberation of an atomic hydrogen from the defect site. While the origin of the defect remains unknown at this time, these results provide a foundation for future potential classification.

It is expected that other sample termination methods, Si wafers, and vacuum systems could potentially lead to additional defects not reported here. While this may mean this is not an exhaustive list, our analysis provides critical insight into the nature of many commonly observed defects. As many groups are now exploring fabrication of larger and more sophisticated atom-scale devices on this substrate, defects interrupting the needed large patterning areas are quickly becoming a limiting factor. Through this now enhanced understanding of the nature of the most common defects, we enable informed refinement of standard H-Si wafer preparation methods, leading to a more reliable platform for the creation of these devices.

**Methods**

Experiments were performed using an Omicron LT STM and an Omicron qPlus LT AFM[76,77] system operating at 4.5 K and ultrahigh vacuum (3 x $10^{-11}$ Torr). STM tips were electrochemically etched from polycrystalline tungsten wire, resistively heated, and field evaporated to clean and sharpen the apex using a field ion microscope (FIM)[78]. AFM tips used the third-generation Giessibl tuning forks with a FIB mounted tungsten wire ($f_0$ ~ 28 kHz, Q-factor ~ 16k-22k)[79]. The tip was cleaned and sharpened in vacuum using a FIM[78]. *In situ* tip conditioning was done by executing controlled contact on a hydrogen-desorbed patch of Silicon[80,81]. Bright contrast AFM tips were formed with controlled contact on both desorbed and H-terminated portions of the surface, while dark contrast AFM tips were formed using only desorbed patches. STHM tips were achieved by creating DBs via tip pulsing until the tip contrast changed to that shown in Fig S1c.

Samples used were highly arsenic-doped (~1.5 x $10^{19}$ atoms/cm$^3$) Si(100). Samples were degassed at 600 °C overnight followed by flash annealing at 1250°C. The samples were then terminated with hydrogen by exposing them to molecular hydrogen ($10^{-6}$ Torr) while the Si sample was held at 330 °C for 2 minutes. The molecular hydrogen was cracked using a tungsten filament held at 1600 °C[82].

Image and data acquisition were done using a Nanonis SPM controller and software, imaging parameters for each of the 6 modes are described in the text. The height setpoint was taken as the tip-sample separation over a H-Si atom with an imaging bias of -1.8V and a current setpoint of 50 pA. The exact magnitude of the Δf(Z) spectroscopy changed between tip shaping events but the general shape and behavior for bright and dark contrast remained consistent throughout multiple tips and tip terminations.

The defect free H-Si Ball and Stick model was the same as in[80]. Defects were manually inserted using Avogadro[83,84]. The geometry of the defect atoms within the lattice were positioned using a Merck Molecular force field (MMFF94)[85] using steepest descent and a convergence of 10e-7. Images of the lattice were colorized and rendered using Mercury[86].

## Acknowledgements

We thank NSERC, and AITF for financial support. We would also like to thank Moe Rashidi and Roshan Achal for valuable discussions regarding the defect origins. We would also like to thank Mark Salomons and Martin Cloutier for their technical expertise. T.D. acknowledges support by NSF-MIP: Platform for the Accelerated Realization, Analysis, and Discovery of Interface Materials (PARADIM, DMR-1539918).

## Author Contributions

J.C. collected all data. Interpretation of all results was done by J.C, T.D., T.H. and R.A.W.. R.A.W. supervised the project. J.C. wrote the manuscript with input from all authors.

## Present address

§Department of Materials Science and Engineering, Cornell University, Ithaca NY 14853, USA.

## Corresponding Author

*wolkow@ualberta.ca

# References

1. Prati, E. & Shinada, T. Atomic scale devices: Advancements and directions. *IEEE Int. Electron Devices Meet.* 1.2.1-1.2.4 (2015). doi:10.1109/IEDM.2014.7046961

2. O'Brien, J. L. *et al.* Towards the fabrication of phosphorus qubits for a silicon quantum computer. *Phys. Rev. B* **64**, 4 (2001).

3. He, Y. *et al.* A two-qubit gate between phosphorus donor electrons in silicon. *Nature* **571**, 371–375 (2019).

4. Fuechsle, M. *et al.* A single-atom transistor. *Nat. Nanotechnol.* **7**, 242–246 (2012).

5. Wyrick, J. *et al.* Atom-by-Atom Fabrication of Single and Few Dopant Quantum Devices. *Adv. Funct. Mater.* 1903475 (2019). doi:10.1002/adfm.201903475

6.  Huff, T. *et al.* Binary atomic silicon logic. *Nat. Electron.* **1**, 636–643 (2018).

7.  Yengui, M., Duverger, E., Sonnet, P. & Riedel, D. A two-dimensional ON/OFF switching device based on anisotropic interactions of atomic quantum dots on Si(100):H. doi:10.1038/s41467-017-02377-4

8.  Kolmer, M. *et al.* Realization of a quantum Hamiltonian Boolean logic gate on the Si(001):H surface. *Nanoscale* **7**, 12325–12330 (2015).

9.  Huff, T. *et al.* Electrostatic Landscape of a Hydrogen-Terminated Silicon Surface Probed by a Moveable Quantum Dot. *ACS Nano.* **13**, 9, 10566-10575 (2019).

10. Rashidi, M. *et al.* Deep Learning-Guided Surface Characterization for Autonomous Hydrogen Lithography. (2019). preprint at arXiv:1902.08818v2

11. Gordon, O. *et al.* Scanning tunneling state recognition with multi-class neural network ensembles. *Rev. Sci. Instrum.* **90**, 103704 (2019).

12. Ziatdinov, M., Fuchs, U., Owen, J. H. G., Randall, J. N. & Kalinin, S. V. Robust multi-scale multi-feature deep learning for atomic and defect identification in Scanning Tunneling Microscopy on H-Si(100) 2x1 surface. (2020).

13. Hamers, R. J. & Köhler, U. K. Determination of the local electronic structure of atomic-sized defects on Si(001) by tunneling spectroscopy. *Cit. J. Vac. Sci. Technol. A* **7**, 2854 (1989).

14. Hossain, M. Z., Yamashita, Y., Mukai, K. & Yoshinobu, J. Model for C defect on Si(100): The dissociative adsorption of a single water molecule on two adjacent dimers. *Phys. Rev. B* **67**, 1–4 (2003).

15. Yu, S.-Y., Kim, H. & Koo, J.-Y. Extrinsic Nature of Point Defects on the Si(001) Surface: Dissociated Water Molecules. *Phys. Rev. Lett.* 100(3):036107 (2018)

16. Warschkow, O. *et al.* Transformation of C-type defects on Si(100)-2x1 surface at room temperature STM/STS study. *Surf. Sci.* **602,** 17, p. 2835-2839 (2008).

17. Sweetman, A., Danza, R., Gangopadhyay, S. & Moriarty, P. Imaging and manipulation of the Si(100) surface by small-amplitude NC-AFM at zero and very low applied bias. *J. Phys. Condens. Matter* **24**, 084009 (2012).

18. Boland, J. J. *Structure of the H-Saturated Si(100) Surface*. **65**, (1990).

19. Boland, J. J. Scanning tunnelling microscopy of the interaction of hydrogen with silicon surfaces. *Adv. Phys.* **42**, 129–171 (1993).

20. Boland, J. J. Role of bond-strain in the chemistry of hydrogen on the Si(100) surface. *Surf. Sci.* **261**, 17–28 (1992).

21. Hamers, R. J. & Wang, Y. Atomically-Resolved Studies of the Chemistry and Bonding at Silicon Surfaces. (1996). doi:10.1021/CR950213K

22. Møller, M. *et al.* Automated extraction of single H atoms with STM: tip state dependency. *Nanotechnology* **28**, 075302 (2017).

23. Sweetman, A. *et al.* Manipulating Si(100) at 5 K using qPlus frequency modulated atomic force microscopy: Role of defects and dynamics in the mechanical switching of atoms. *Phys. Rev. B* **84**, 85426 (2011).

24. Kuzmin, M. *et al.* Imaging empty states on the Ge(100) surface at 12K. *Phys. Rev. B* **98**, 1–10

(2018).

25. Weiss, C. *et al.* Imaging Pauli Repulsion in Scanning Tunneling Microscopy. *Phys. Rev. Lett.* **105**, 086103 (2010).

26. Gross, L., Mohn, F., Moll, N., Liljeroth, P. & Meyer, G. The Chemical Structure of a Molecule Resolved by Atomic Force Microscopy. *Science (80-. ).* **325**, 1110–4 (2009).

27. Mönig, H. *et al.* Understanding Scanning Tunneling Microscopy Contrast Mechanisms on Metal Oxides: A Case Study. *ACS Nano* **7**, 10233–10244 (2013).

28. Mönig, H. *et al.* Quantitative assessment of intermolecular interactions by atomic force microscopy imaging using copper oxide tips. *Nat. Nanotechnol.* **13**, 371–375 (2018).

29. Mohn, F., Schuler, B., Gross, L. & Meyer, G. Different tips for high-resolution atomic force microscopy and scanning tunneling microscopy of single molecules. *Appl. Phys. Lett.* **102**, 1–5 (2013).

30. Kichin, G., Weiss, C., Wagner, C., Stefan Tautz, F. & Temirov, R. Single Molecule and Single Atom Sensors for Atomic Resolution Imaging of Chemically Complex Surfaces. *J. Am. Chem. Soc* **133**, 16847–16851 (2011).

31. Temirov, R., Soubatch, S., Neucheva, O., Lassise, A. C. & Tautz, F. S. A novel method achieving ultra-high geometrical resolution in scanning tunnelling microscopy. *New J. Phys.* **10**, 053012 (2008).

32. Simic-Milosevic, V., Mehlhorn, M. & Morgenstern, K. Imaging the Bonds of Dehalogenated Benzene Radicals on Cu(111) and Au(111). *ChemPhysChem* **17**, 2679–2685 (2016).

33. Huff, T. R. *et al.* Atomic White-Out: Enabling Atomic Circuitry through Mechanically Induced Bonding of Single Hydrogen Atoms to a Silicon Surface. *ACS Nano* **11**, 8636–8642 (2017).

34. Achal, R. *et al.* Lithography for robust and editable atomic-scale silicon devices and memories. *Nat. Commun.* **9**, 1–8 (2018).

35. Martínez, J. I., Abad, E., González, C., Flores, F. & Ortega, J. Improvement of Scanning Tunneling Microscopy Resolution with H-Sensitized Tips. (2012). doi:10.1103/PhysRevLett.108.246102

36. Sharp, P. *et al.* Identifying passivated dynamic force microscopy tips on H:Si(100). *Appl. Phys. Lett.* **100**, 233120 (2012).

37. Pérez, R., Payne, M. C., Štich, I. & Terakura, K. Role of Covalent Tip-Surface Interactions in Noncontact Atomic Force Microscopy on Reactive Surfaces. *Phys. Rev. Lett.* **78**, 678–681 (1997).

38. Sugimoto, Y. & Onoda, J. Force spectroscopy using a quartz length-extension resonator. *Appl. Phys. Lett.* **115**, (2019).

39. Bellec, A. *et al.* Electronic properties of the n-doped hydrogenated silicon (100) surface and dehydrogenated structures at 5 K. *Phys. Rev. B* **80**, (2009).

40. Engelund, M. *et al.* The butterfly-a well-defined constant-current topography pattern on Si(001):H and Ge(001):H resulting from current-induced defect fluctuations †. *Phys. Chem. Chem. Phys* **18**, 19309–19317 (2016).

41. Lyding, J. W., Shen, T.-C., Abeln, G. C., Wang, C. & Tucker, J. R. Nanoscale patterning and selective chemistry of silicon surfaces by ultrahigh-vacuum scanning tunneling microscopy. *Nanotechnology* **7**, 128 (1996).

42. Schofield, S. R. *et al.* Quantum engineering at the silicon surface using dangling bonds. *Nat. Commun.* **4**, 1–7 (2013).

43. Taucer, M. *et al.* Single-electron dynamics of an atomic silicon quantum dot on the H-Si (100)-(2x1) surface. *Phys. Rev. Lett.* **112**, 1–5 (2014).

44. Labidi, H. *et al.* Scanning tunneling spectroscopy reveals a silicon dangling bond charge state transition. *New J. Phys.* **17**, (2015).

45. Sinthiptharakoon, K. *et al.* Investigating individual arsenic dopant atoms in silicon using low-temperature scanning tunnelling microscopy. *J. Phys. Condens. Matter J. Phys. Condens. Matter* **26**, 1–8 (2014).

46. Liu, L., Yu, J. & Lyding, J. W. Atom-resolved three-dimensional mapping of boron dopants in Si(100) by scanning tunneling microscopy. *Cit. Appl. Phys. Lett* **78**, 386 (2001).

47. Liu, L., Yu, J. & Lyding, J. W. Subsurface dopant-induced features on the Si(100)2x1:H surface: fundamental study and applications. *IEEE Trans. Nanotechnol.* **1**, 176–183 (2002).

48. Domke, C., Ebert, P., Heinrich, M. & Urban, K. Microscopic identification of the compensation mechanism in Si-doped GaAs. *Phys. Rev. B* **54**, (1996).

49. Lee, D. H. & Gupta, J. A. Tunable field control over the binding energy of single dopants by a charged vacancy in GaAs. *Science* **330**, 1807–10 (2010).

50. Ebert, P. *et al.* Thermal formation of Zn-dopant-vacancy complexes on InP(110) surfaces. *Phys. Rev. B* **53**, (1996).

51. Kajiyama, H. *et al.* Room-Temperature Adsorption of Si Atoms on H-terminated Si(001)-2 x 1 Surface. *J. Phys. Soc. Japan* **74**, 389–392 (2005).

52. Wang, C., Zhang, Y. & Jia, Y. A new Si tetramer structure on Si (001). *Solid State Sci.* **11**, 1661–1665 (2009).

53. Bellec, A., Riedel, D., Dujardin, G., Rompotis, N. & Kantorovich, L. N. Dihydride dimer structures on the Si(100):H surface studied by low-temperature scanning tunneling microscopy. *Phys. Rev. B* **78**, (2008).

54. Suwa, Y. *et al.* First Principles Study of Dihydride Chains on H-Terminated Si(100)-2x1 Surface. *Jpn. J. Appl. Phys.* **45**, 2200–2203 (2006).

55. Suwa, Y. *et al.* Formation of dihydride chains on H-terminated Si ( 100 ) – 2 × 1 surfaces: Scanning tunneling microscopy and first-principles calculations. *Phys. Rev. B* **74**, 205308 (2006).

56. Suwa, Y. *et al.* Dopant-Pair Structures Segregated on a Hydrogen-Terminated Si(100) Surface. *Phys. Rev. Lett.* **90**, (2003).

57. Cheng, C. C. & Yates, J. T. H-induced surface restructuring on Si(100): Formation of higher hydrides. *Phys. Rev. B* **43**, 4041–4045 (1991).

58. Chen, D. & Boland, J. J. Spontaneous roughening and vacancy dynamics on Si(100)-2×1:Cl. *Surf. Sci.* **518**, L583–L587 (2002).

59. Weakliem, P. C., Zhang, Z. & Metiu, H. Missing dimer vacancies ordering on the Si(100) surface. *Surf. Sci.* **336**, 303–313 (1995).

60. Owen, J. H. G., Bowler, D. R., Goringe, C. M., Miki, K. & Briggs, G. A. D. Identification of the Si(001) missing dimer defect structure by low bias voltage STM and LDA modelling. *Surf. Sci.* **341**, 1042–

1047 (1995).

61. *Springer-Verlag Berlin Heidelberg GmbH*. *Noncontact Atomic Force Microscopy* (Springer-Verlag, 1999). doi:10.1007/978-3-662-06431-3

62. Chabal, Y. J. & Raghavachari, K. New Ordered Structure for the H-Saturated Si(100) Surface: The (3 x 1) Phase. *Phys. Rev. Lett.* **54**, 1055–1058 (1985).

63. Maeng, J. Y., Kim, S., Jo, S. K., Fitts, W. P. & White, J. M. Absorption of gas-phase atomic hydrogen by Si(100): Effect of surface atomic structures. *Cit. Appl. Phys. Lett* **79**, 36 (2001).

64. Buehler, E. J. & Boland, J. J. *Surface Science Letters Identification and characterization of a novel silicon hydride species on the Si(100) surface*. *Surface Science* **425**, (1999).

65. Hitosugi, T. *et al.* Direct Observation of One-Dimensional Ga-Atom Migration on a Si(100)-(2x1)-H Surface: A Local Probe of Adsorption Energy Variation. *Phys. Rev. Lett.* **83**, (1999).

66. Ballard, J. B. *et al.* Spurious dangling bond formation during atomically precise hydrogen depassivation lithography on Si(100): The role of liberated hydrogen. *J. Vac. Sci. Technol. B, Nanotechnol. Microelectron. Mater. Process. Meas. Phenom.* **32**, 21805 (2014).

67. Trenhaile, B. R., Agrawal, A. & Weaver, J. H. Oxygen atoms on Si(100)-(2x1): Imaging with scanning tunneling microscopy. *Appl. Phys. Lett.* **89**, 173118 (2006).

68. Hossain, M. Z., Yamashita, Y., Mukai, K. & Yoshinobu, J. Model for C defect on Si"100…: The dissociative adsorption of a single water molecule on two adjacent dimers. doi:10.1103/PhysRevB.67.153307

69. Sobotík, P. & Ošt'ádal, I. Transformations of C-type defects on Si(100)-2 x 1 surface at room temperature-STM/STS study. *Surf. Sci.* **602**, 2835–2839 (2008).

70. Roberson, M. A. & Estreicher, S. K. *Vacancy and vacancy-hydrogen complexes in silicon*. **49**, (1994).

71. Xu, H. Electronic structure of hydrogen-vacancy complexes in crystalline silicon: A theoretical study. *Phys. Rev. B* **46**, 1403–1422 (1992).

72. Zundel, T. & Weber, J. Trap-limited hydrogen diffusion in boron-doped silicon. *Phys. Rev. B* **46**, (1992).

73. Herrero, C. P., Stutzmann, M. & Breitschwerdt, A. Boron-hydrogen complexes in crystalline silicon. *Phys. Rev. B* **43**, 1555–1575 (1991).

74. Ong, C. K. & Khoo, G. S. The nature of the hydrogen-boron complex of crystalline Si. *J. Phys. Condens. Matter* **3**, 675–680 (1991).

75. Rizk, R., de Mierry, P., Song, C., Ballutaud, D. & Pajot, B. Spectroscopic investigation of hydrogen-dopant complexes in bulk *p*-type and implanted *n*-type crystalline silicon. *J. Appl. Phys.* **70**, 3802–3807 (1991).

76. Giessibl, F. J. High-speed force sensor for force microscopy and profilometry utilizing a quartz tuning fork. *Appl. Phys. Lett.* **73**, 3956–3958 (1998).

77. Giessibl, F. J. The qPlus sensor, a powerful core for the atomic force microscope. *Rev. Sci. Instrum.* **90**, (2019).

78. Rezeq, M., Pitters, J. & Wolkow, R. Tungsten nanotip fabrication by spatially controlled field-assisted reaction with nitrogen. *J. Chem. Phys.* **124**, (2006).

79. Labidi, H. *et al.* New fabrication technique for highly sensitive qPlus sensor with well-defined spring constant. *Ultramicroscopy* **158**, 33–37 (2015).

80. Labidi, H. *et al.* Indications of chemical bond contrast in AFM images of a hydrogen-terminated silicon surface. *Nat. Commun.* **8**, (2017).

81. Rashidi, M. & Wolkow, R. A. Autonomous Scanning Probe Microscopy in situ Tip Conditioning through Machine Learning. *ACS Nano* **12**, 5185–5189 (2018).

82. Pitters, J. L., Piva, P. G. & Wolkow, R. A. Dopant depletion in the near surface region of thermally prepared silicon (100) in UHV. *J. Vac. Sci. Technol. B* **30**, 021806+ (2012).

83. Avogadro: an open-source molecular builder and visualization tool.

84. Hanwell, M. D. *et al.* Avogadro: an advanced semantic chemical editor, visualization, and analysis platform. *J. Cheminform.* **4**, 17 (2012).

85. Halgren, T. A. Merck molecular force field. I. Basis, form, scope, parameterization, and performance of MMFF94. *J. Comput. Chem.* **17**, 490–519 (1996).

86. Macrae, C. F. *et al.* Mercury CSD 2.0 – new features for the visualization and investigation of crystal structures. *J. Appl. Crystallogr.* **41**, 466–470 (2008).